\documentclass[conference]{IEEEtran}
\IEEEoverridecommandlockouts
\usepackage{cite}
\usepackage{amsmath,amssymb,amsfonts}
\usepackage{algorithm}
\usepackage{algpseudocode}
\usepackage{amsfonts}
\usepackage{graphicx}
\usepackage{textcomp}
\usepackage{xcolor}
\def\BibTeX{{\rm B\kern-.05em{\sc i\kern-.025em b}\kern-.08em
    T\kern-.1667em\lower.7ex\hbox{E}\kern-.125emX}}
\begin{document}

\title{Targeted Hardening of Electric Distribution System  for Enhanced Resilience against Earthquakes\\
}

\author{
\IEEEauthorblockN{Mahan Fakouri Fard$^\dag$, Mostafa Sahraei-Ardakani$^\dag$, Ge Ou$^\ddag$, and Mingxi Liu$^\dag$}
\IEEEauthorblockA{$^\dag$Department of Electrical and Computer Engineering~$^\ddag$Department of Civil and Environmental Engineering\\
University of Utah\\
Salt Lake City, UT, USA, 84112\\
\{mahan.fakourifard, mostafa.ardakani, ge.ou, mingxi.liu\}@utah.edu}
}

\maketitle

\begin{abstract}
Securing the power system from catastrophic natural disasters is a rising problem in power system operation and planning. This paper particularly considers earthquake and aims to evaluate and improve the resilience of power distribution networks by developing a novel hardware hardening framework. In the proposed framework, fragility curves of the network equipment are used to represent equipment failure probabilities when facing an earthquake, and failure scenarios of the distribution network are obtained via the Monte Carlo method. Based on the distribution network topology and the locations of essential loads, various hardware hardening strategies are determined within the proposed framework. Through a series of resilience and economic analyses, the optimal hardening strategy is determined to improve the resilience and supply the essential loads during and after the earthquake. The efficacy of the proposed approach is examined through simulations on an IEEE 33-bus test feeder.
\end{abstract}


\section{Introduction}
Natural disasters such as earthquakes are one of the main causes of widespread power system blackouts. The increasing number of such incidents in recent years and the growing dependence of the people’s life and the nation’s critical infrastructure on the electric power network have attracted growing attention to power system resilience enhancement \cite{Panteli_theGrid_2015}. Each single element of the power system can be affected by a natural disaster in different ways and each natural disaster has its own attributes and destruction forces, leading to the complexity of power system resilience enhancement.

Most existing studies focus on improving the resilience of the transmission networks \cite{Aziz_transmissionResilience_2021}, However, distribution network resilience is equally important as it concerns urban areas where the impacts from natural disasters are more devastating and can cause further casualties. As many of the critical infrastructures, e.g., hospitals and military bases, rely on electricity provided by the distribution network, maintaining uninterrupted power supply to the critical infrastructures is demanded during a natural disaster. The power grid can be affected by different types of natural disasters. Most existing works on power system resilience focus on floods \cite{Costa_flood_2017} and hurricanes \cite{Shen_hurricane_2021} while studies focusing on earthquakes are limited. Comparing with other natural disasters, earthquake happens in a very short time, is hard to predict, and can simultaneously impact the whole distribution network. This paper is dedicated to investigating the resilience of distribution network against earthquakes.

Power system resilience enhancement measures are normally categorized into hardening-oriented and operation-oriented. Reinforcing poles and substations as well as vegetation management are examples of the former \cite{panteli_2017}, while rescheduling generation, strategic microgrid islanding, and priority based load shedding are practical measures of the latter \cite{ESPINOZA_2016}. Operation-oriented resilience measures are post-event strategies that focus on recovering the network operation after the damages are made. Though we do not discount the merits of operation-oriented strategies, strengthening the network hardware, on the other hand, is a pre-event strategy that can efficiently prevent high-impact casualties and reduce economical burdens. However, studies on hardware hardening measures for distribution network is quite limited. Among the limited related works, a preventive preparedness for energy-carrying microgrids to cope with severe weather events through allocating pre-storm resources was developed in \cite{amirioun_2019}. A new criterion was introduced to measure the preparedness of microgrids against storms and a multi-objective optimization problem was formulated to minimize the amount of load reduction and meet this criterion.  Though \cite{amirioun_2019} presents a good implementation of proactive management for resilience enhancement, it is not effective in handling earthquakes as disaster forecasting is not available for resource allocation.

Manshadi \emph{et al.} \cite{Manshadi_2015} devised an exponential index to evaluate the resilience of multi-energy-carrying microgrids in the face of destruction of power distribution lines, natural gas distribution lines, and production resources. The index represents the extra cost of microgrid operation after subversion and can be used by investors and power system beneficiaries to evaluate the economic loss of the incident. However, this index fails to provide technical information about the temporal behaviors of the system during or after the incident. 

Existing hardware hardening studies fall short in handling distribution network resilience in the face of earthquake and the economical aspect as a deciding factor in the hardening procedure is rarely considered in the literature. In this paper, we develop a novel hardware hardening framework to protect the essential loads in the distribution network during earthquake that was not considered in the literature. The contribution of this paper is three-fold: 1) We significantly improve the distribution network resilience evaluation mechanism by introducing three normalized indices to evaluate the resilience during and after the earthquake; 2) A comprehensive decision making framework is, for the first time, developed to determine the optimal hardware hardening strategy that comprehensively considers the impact of an earthquake on the distribution network as well as the costs and benefits; and 3) The proposed method can be readily customized to achieve different operation objectives and extended to handle different natural disasters.

\section{Methodology}
\subsection{Resilience Evaluation Index}\label{Sec_evaluation}
\begin{figure}[!htb]
  \includegraphics[width=\linewidth]{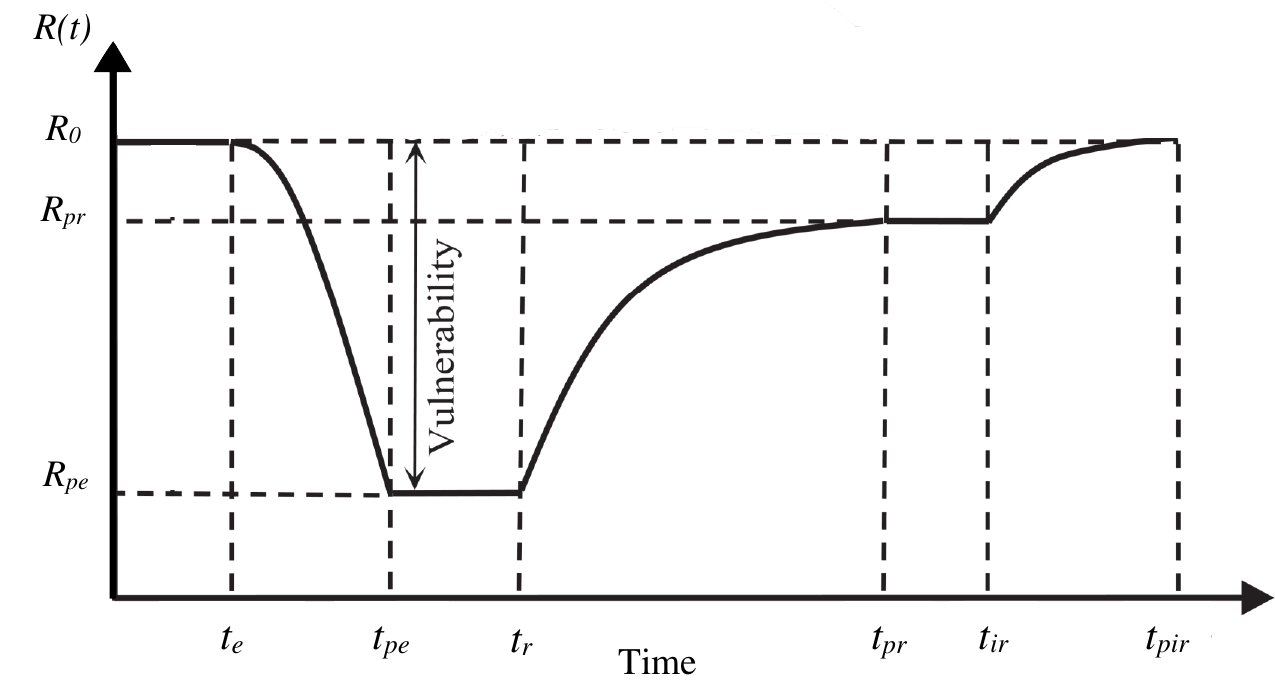}
  \caption{A generic network resilience curve associated with an event.}
  \label{resilience curve}
\end{figure}

In this section, we propose a new system performance index to represent the distribution network resilience and introduce three resilience indices. Fig. \ref{resilience curve} presents a generic network resilience curve. 
We first propose a performance index $R(t)$ concerning the average amount of power supplied at time $t$ as
\begin{equation}\label{performance index}
    R(t)={L_C(t)}/{L_T(t)},
\end{equation}
where $L_C(t)$ is the sum of the supplied essential loads $L_C^{\Omega}(t)$ and common loads $L_C^C(t)$ of the network, and $L_T(t)$ is the sum of the total loads in the grid, both supplied and unmet. Note that, $R(t)=1$ in the case of uninterrupted delivery, and $R(t)=0$ in the case of complete outage, indicating the loss of all loads of the network.

As shown in  Fig. \ref{resilience curve}, after an event occurs at time $t_e$ and in the post event degraded state time interval $[t_{pe},t_r]$, the supplied load is significantly reduced due to the failure of network equipment.
To determine the resilience criteria, let $R_0=R(t_e)$, $R_{pe}=R(t_{pe})$, and $R_{pr}=R(t_{pr})$ be the system performance indices before the event, after the event and network failure, and after the network recovery from the event, respectively. Consequently, we consider the interval $[t_{pe},t_r]$ as the system degraded phase, where $t_r$ is the time when the recovery begins. We define the normalized vulnerability index $\zeta_V$ as
\begin{equation}\label{VI}
    \zeta_V=({R_0-R_{pe}})/{R_0},
\end{equation}
whose value is $0$ for no degraded state and $1$ for complete blackout. This vulnerability index determines the decline ratio of the average supplied load after and before the event occurrence.
To see the impact of time, we introduce the degradation index as
\begin{equation}\label{DI}
    \zeta_D={\int_{t_e}^{t_{pe}}(R_0-R(t))dt}/({R_0 (t_{pe}-t_e)}),
\end{equation}
which takes the value of $1$ for a sudden complete blackout that lasts until recovery begins, and $0$ for no lost load. 
In addition to $\zeta_V$ and $\zeta_D$, we introduce another normalized index $\zeta_E$ that evaluates the amount of energy not provided due to the event as
\begin{equation}\label{UEI}
    \zeta_E={\int_{t_e}^{t_{pir}}(R_0-R(t))dt}/){R_0 (t_{pir}-t_e)}),
\end{equation}
which indicates the rate of performance decrease over the entire event period. $\zeta_E$ assumes the value of $0$ in the ideal state without a drop, and assumes the value of $1$ in the worst-case scenario, indicating the maximum drop in the fastest time.
\subsection{Hardware Hardening Strategy Generation Framework}
\subsubsection{Phase I - Determine network goal \& Data Collection}\label{p1s1}
Each distribution network is designed and operated based on specific contingency standards, e.g., limit for maximum duration of degraded state. These standards and goals should be determined by the network operators and planners. The main goal of this paper is to maximize the amount of essential load provided during the earthquake and the improvement of the resilience indices, i.e., $\zeta_V$, $\zeta_D$, and $\zeta_E$. The available resources for hardware reinforcement will also be considered. The event scenario will be selected in this step.

A network graph will be generated where buses are represented as nodes and power lines are represented as branches. General specifications of network equipment such as electric poles and wires in each line, types of lines, generation and load information, and locations and values of the essential loads are collected. Fragility curves based on the event of all equipment will also be collected. Each fragility curve presents the failing possibility of each equipment, hardened and not hardened, in different intensities of the accident.
\subsubsection{Phase II - Modeling}
\paragraph{Step 1 - Model line failure}\label{p2s1}
 In a distribution network, each line consists of several poles, wires, and supporting arms, each of which has a probability of failure. The calculation of line failure probability depends on the line type. For an overhead line, the number of poles it crosses and the fragility curve of each pole must be used; for the underground cables, the failure probability will be driven by the fragility curves directly, which are assumed to be uniform for the entire length of this line. Uniformly,  the failing probability $\mathbb{P}_{\ell}$ of line $\ell$ can be calculated as
\begin{equation}\label{totall prob}
    \mathbb{P}_{\ell}=1-\overline{\mathbb{P}}_{\ell},
\end{equation}
where $\overline{\mathbb{P}}_{\ell}=\prod_{k\in \mathcal{K}_{\ell}}(1-\mathbb{P}_{\ell,k})$, $\mathbb{P}_{\ell,k}=\mathbb{P}_{\ell,k}^{w}+\mathbb{P}_{\ell,k}^{p}-\mathbb{P}_{\ell,k}^{w}\mathbb{P}_{\ell,k}^{p}$, $\mathcal{K}_{\ell}$ is the set of available poles crossed by line $\ell$, $\mathbb{P}_{\ell,k}^{w}$ and $\mathbb{P}_{\ell,k}^{p}$ are the failure probabilities of the wire and the pole, respectively, and ${\overline{\mathbb{P}}}_{\ell}$ is the probability of not failing.
\paragraph{Step 2 - Evaluate network resilience}\label{p2s2}
This step leverages the Monte Carlo algorithm \cite{Reiter_MonteCarlo_2007} to determine the lines disconnected from the network during an earthquake \cite{panteli_2017}. Algorithm \ref{Monte_Carlo_Alg} presents the procedure of network resilience evaluation. \begin{algorithm} 
\caption{Network resilience evaluation}\label{Steps for evaluation of the network resilience.}
\begin{algorithmic}[1]
\State Set the maximum number of Monte Carlo iterations to $N$ and initialize the iteration with $n=1$
\While{$n\leq N$} 
\State Determine the failed lines
\State Collect and save supplied essential and common loads
\State Start the recovery phase. Recover two failed lines per hour until full restoration
\EndWhile
\State Calculate the average supplied essential and common load at each time using \eqref{calculate supplied load}-\eqref{total supplied load} 
\State Draw the resilience curve
\end{algorithmic}
\label{Monte_Carlo_Alg}
\end{algorithm} In each iteration of Algorithm \ref{Monte_Carlo_Alg}, lines that fail as a result of the earthquake are identified. When a line goes out of service, its representative branch is removed from the network and the network graph is re-drawn. Then, based on the new network graph, the amount of supplied essential and common loads are calculated with performing load dispatch. In the load dispatch, we omit the reactive power and line losses and assume unlimited line capacity. To calculate the amount of load supplied, connections between the loads and the power generating buses must be checked -- a disconnected load cannot be supplied. Note that after the accident occurs, the connectivity of the network graph may change due to the line cuts, thus the new graph may consists of several separate sub-graphs. Let $\mathcal{G}$ be the set of sub-graphs in each Monte Carlo iteration graph, then the total amount of load supplied in each internally connected sub-graph $\mathcal{G}_i$ is calculated as
 \begin{equation} \label{calculate supplied load}
    L_C^{\mathcal{G}_i}(t)=\min\left\{ \sum_{j\in \mathcal{G}_i}P^g_j(t), \sum_{j\in \mathcal{G}_i}P^l_j(t) \right\},
\end{equation}
where $P_j^g(t)$ and $P_j^l(t)$ are the amount of generation and load for node $j$, respectively. Consequently, the total supplied load in each iteration is
\begin{equation}\label{total supplied load}
    L_C(t)=\sum_{i=1}^{|\mathcal{G}|}L_C^{\mathcal{G}_i}(t),
\end{equation}
where $|\mathcal{G}|$ is the cardinality of $\mathcal{G}$ denoting the total number of sub-graphs. At the end of the Monte Carlo iterations, the average amount of supplied load at each time interval is calculated using \eqref{performance index}. Plotting the average supplied load over the time provides the network resilience curve as shown in Fig. \ref{resilience curve}.
\paragraph{Step 3 - Determine hardening strategies}\label{p2s3}
Depending on the locations of the essential loads and the network structure, various strategies can be proposed to harden the network lines and cables. Each strategy is a set of lines, each of which connects the essential loads and the available power resources in the network. To find these strategies, we rely on a path finding algorithm -- Dijkstra’s algorithm \cite{PENG_djikestra_2012}. Inputs of this algorithm are the network graph and the locations of generation and essential loads, while the outputs are the available paths that connect each essential load to a capable generation source. This algorithm will be executed for each essential load to determine the paths between that essential load and generation sources. Each combination of the paths (lines) generated for all essential loads is a candidate strategy $s$ in the strategy set $\mathcal{S}$.
\paragraph{Step 4 - Cost benefit analysis}\label{p2s4}
The cost of each strategy determined in Step 3 consists of the cost of hardware hardening for fortifying poles, retaining arms of each line, strengthening the underground cable ducts, and strengthening the network posts and feeders. For the $s$th strategy in the strategy set $\mathcal{S}$, its cost $C(s)$ is calculated as
\begin{equation} \label{cost}
    C(s)=\sum_{\ell\in s}\left[\xi_\ell c_{\ell}^o |\mathcal{K}_\ell| +(1-\xi_\ell ) c_{\ell}^u  L_\ell \right]+C_0,
\end{equation}
where $\xi_\ell$ is the line type that being $1$ indicates an overhead line and 0 for an underground cable, $c_{\ell}^o$ and $c_{\ell}^u$ are the hardening cost of the overhead line and underground cable, respectively, cardinality $|\mathcal{K}_\ell|$ is the number of poles crossed by line $\ell$, $L_\ell$ is the length of the underground cable $\ell$, and $C_0$ is the base hardening cost to fortify the distribution substation.

The benefit of each strategy consists of various elements with the largest portion of it being the value of the essential loads that will be protected from interruption. Valuation for essential loads depends on the nature of the loads and it will be done by the planners and owners of the network based on asset management and importance measure analysis \cite{Fang_importanceMeasure_2016}. Another element within the benefit is the savings from not replacing or rebuilding poles or cables. However, note that a pole that is severely damaged in an event and needs to be replaced or overhauled will impose certain cost on the network and that cost can be prevented by strengthening the hardware before the event. In fact, benefit is the monetary value that the network operator would have had to pay if the hardening strategy had not been implemented. Thus, the benefit $B(s)$ of the $s$th strategy can be calculated as
\begin{equation}\label{benefit}
\begin{aligned}
    B(s)=\sum_{i\in \Omega}(L_{C,i}\omega_i)+L_C^C\bar{\omega}+\Gamma \sum_{j\in J(s)}\chi_j (\mathbb{P}_j-\mathbb{P}_j^H),
\end{aligned}
\end{equation}
where $\Omega$ is the set of essential loads, $L_{C,i}$ and $L_C^C$ are the amount of the supplied load of essential load $i$ and the common load in $[t_{pe},t_r]$, respectively, $\omega_i$ and $\bar{\omega}$ are the value of essential load $i$ and the common load, respectively, $\Gamma$ is the return ratio of the event, $J(s)\in \{\mathcal{K_\ell}|\ell\in s\}$ is the set of contributing poles in the $s$th strategy, $\chi_j$ is the repair cost of the $j$th pole, and $\mathbb{P}_j^H$ and $\mathbb{P}_j$ are the failure probabilities of pole $j$ in hardened and not hardened modes, respectively.

Another element within the benefit analysis is the rate of improvement of the resilience indices in \eqref{VI}, \eqref{DI}, and \eqref{UEI}. The resilience improvement $\lambda(s)$ of the $s$th strategy is defined as
\begin{equation} \label{resilience factor}
\begin{aligned}
\lambda(s)=(\zeta_V-\zeta_V^s)+(\zeta_D-\zeta_D^s)+(\zeta_E-\zeta_E^s),
\end{aligned}
\end{equation}
where $(\zeta_V^s$,$\zeta_D^s$,$\zeta_E^s)$ are the resilience indices after implementing the $s$th strategy. Once the cost and benefit are calculated, benefit-to-cost ratio $\alpha(s)$ will be calculated as
\begin{equation} \label{cost to benefit}
\begin{aligned}
\alpha(s)={B(s)(1+\lambda(s))}/{C(s)} .
\end{aligned}
\end{equation}
 All strategies will be listed in descending order of $\alpha(s)$.
 \paragraph{Step 5 - Determine the optimal strategy}\label{p2s5}
The ``optimal'' strategy is apparently at the top of the strategy list. However, it is not yet clear whether this strategy can meet the network goals determined in Section \ref{p1s1}. With the implementation of the first strategy on the list, the probabilities of failure of the overhead lines, poles crossed by the overhead lines, and the underground cables are reduced. The network resilience status of this strategy should be evaluated and compared with the target criteria. If the target criteria are met, this strategy is selected as the optimal one; if it fails to meet the target criteria, the next strategy on the list will be evaluated. Note that the cost of each strategy should be no more than the budget. The case where no strategy reaches the target criteria indicates that the criteria cannot be achieved under the specified accident scenario with the specified budget, and the target criteria should be modified. The detailed procedure of determining the optimal hardening strategy is shown in Algorithm \ref{Proposed algorithm}.
\begin{algorithm}[!htb]
\caption{Optimal hardware hardening strategy determination}\label{Proposed algorithm}
\begin{algorithmic}[1]
\State Determine the event scenario and network goals
\State Determine the essential load data and budget
\State Calculate failure probabilities of lines based on fragility curves
\State Evaluate resilience indices $\zeta_V$, $\zeta_D$, and $\zeta_E$
\If{$\zeta_V$, $\zeta_D$, and $\zeta_E$ meet the target criteria}
\State No hardening is needed
\Else
    \State Determine all possible hardening strategies by following Section \ref{p2s3} and evaluate the corresponding network resilience indices
    \State Conduct cost-benefit analysis of each strategy $s \in \mathcal{S}$
    \State Sort the strategies based on the benefit-to-cost ratio
    \State Initialize strategy indicator $s=1$
    \While{$s\leq|\mathcal{S}|$}
    \If {$\zeta_V^s$, $\zeta_D^s$, and $\zeta_E^s$ meet the target criteria}
    \State Output strategy $s$ as the optimal strategy
    \Else
    \State $s=s+1$
        \EndIf
    \EndWhile
    \If{$s>|\mathcal{S}|$}    
        \State Hardening is not affordable for the selected event or the provided budget
        \State Modify the target criteria
\State Return to Line 5
\EndIf
\EndIf
\end{algorithmic}
\end{algorithm}
 Note that Algorithm \ref{Proposed algorithm} is agnostic to specific conditions, thus can be adopted for all natural disasters, distribution networks, and transmission networks.
 \section{Case Study}
In this paper, we consider an earthquake with an averaged peak ground acceleration (PGA) of 0.5 $g$ across the region where the distribution network equipment is located. The studied event is equivalent to the 2010 Haiti earthquake \cite{Athena_Heiti_2010}. We implement the proposed framework to the standard IEEE 33-bus test feeder \cite{Baran_Network_1989}. As shown in Fig. \ref{Graph of the network in one of the Monte Carlo iterations without hardening}, in this distribution feeder,  four distributed energy resources (DER) are connected to buses 6, 12, 18, and 29, and four essential loads are connected to buses 9, 17, 19, and 25. The generation capacity $P_G^{Max}$ and common load data can be found in TABLE \ref{bus,gen,load data}. The value of common loads $\bar{\omega}$ is \$10/kW.\begin{table}[!htb]
\centering
\caption{\label{bus,gen,load data}Bus, Generation and Load Data}
 \begin{tabular}{ccc|ccc} 
 \hline
 Bus & $P_G^{Max}$ & Load & Bus & $P_G^{Max}$ & Load \\ 
 \hline
 1 & 700 kW & 0 & 5,7,11,13-15 & 0 & 35 kW  \\ 
 2,8-10 & 0 & 25 kW & 16,17,19,20,30,31 & 0 & 20 kW    \\
 3 & 0 & 15 kW &  18,29 & 150 kW & 0   \\
 4 & 0 & 45 kW & 21,22,26,32,33 & 0 & 40 kW  \\
 6,12 & 100 kW & 0 & 23-25,27,28 & 0 & 30 kW\\
 \hline
 \end{tabular}
\end{table} 
 The essential load data are presented in TABLE \ref{essential loads}. \begin{table}[!htb] 
\centering
\caption{\label{essential loads}Essential Load Data}
\begin{tabular}{c c c c}
\hline
    Essential load & $\omega$ (\$/kW) & Load Amount (kW)& Connection Bus \\
     \hline
     $\Omega_1$ & 1,200 & 80 & 9 \\
     $\Omega_2$ & 1,000 & 60 & 25 \\
     $\Omega_3$ & 1,400 & 60 & 19 \\
     $\Omega_4$ & 1,600 & 45 & 17 \\
     \hline
\end{tabular}
\end{table}
The fragility curves of the poles and underground cables presented in \cite{Xinzhenglu_fragilitycurves_2020} are used in our simulations. The cost data of power pole, underground cable, and substation hardening are extracted from \cite{Guidelines_electricityauthority_2018}. Specifically, the costs of hardening a power pole $c_\ell^o$, the underground cable $c_\ell^u$, and the distribution substation $C_0$ are \$500/pole, \$4,000/mile, and \$10,000, respectively.
Table \ref{line data} presents four groups of network lines used in the simulation. \begin{table}[!htb]
\centering
\caption{\label{line data} 33-bus Network Lines Data }
\begin{tabular}{c c c c c c}
\hline
    Lines & Type & $|\mathcal{K}_{\ell}|$ or $|L_{\ell}|$& $\chi$ & $\mathbb{P}$& $\mathbb{P}^H$ \\
\hline    
     1 - 17 & Overhead & 20 & \$1,200  & 0.88 & 0.03\\
     18 - 25 & Cable & 1.2 mile & \$10,000 & 0.9 & 0.02\\
     26 - 32 & Overhead & 15 & \$1,000 & 0.74 & 0.05\\
     33 - 36 & Overhead & 25 & \$1,500 & 0.8 & 0.1\\
     \hline
\end{tabular}
\end{table} As shown in Table \ref{line data}, the network consists of one underground cable group and three groups of overhead lines with different numbers of poles, failure probability without hardening $\mathbb{P}$, and hardened modes $\mathbb{P}^H$. It is assumed that lines in the same group are the same in terms of lifetime, conditions of construction, and maintenance. The failure probabilities (hardened and not hardened) are obtained by calculating ${\overline{\mathbb{P}}}$ and the fragility curves of each equipment. In an earthquake with an anverage PGA of 0.5 $g$, the failure probabilities of the pre-hardening and post-hardening wires are 10\% and 1\%, respectively \cite{Xinzhenglu_fragilitycurves_2020}. The return period of the earthquake is set at 475 years, i.e., corresponding to 10 \% probability of exceedance in 50 years.
\subsection{Resilience Evaluation before Hardening}
\begin{figure}
    \centering
    \includegraphics[width=0.75\linewidth]{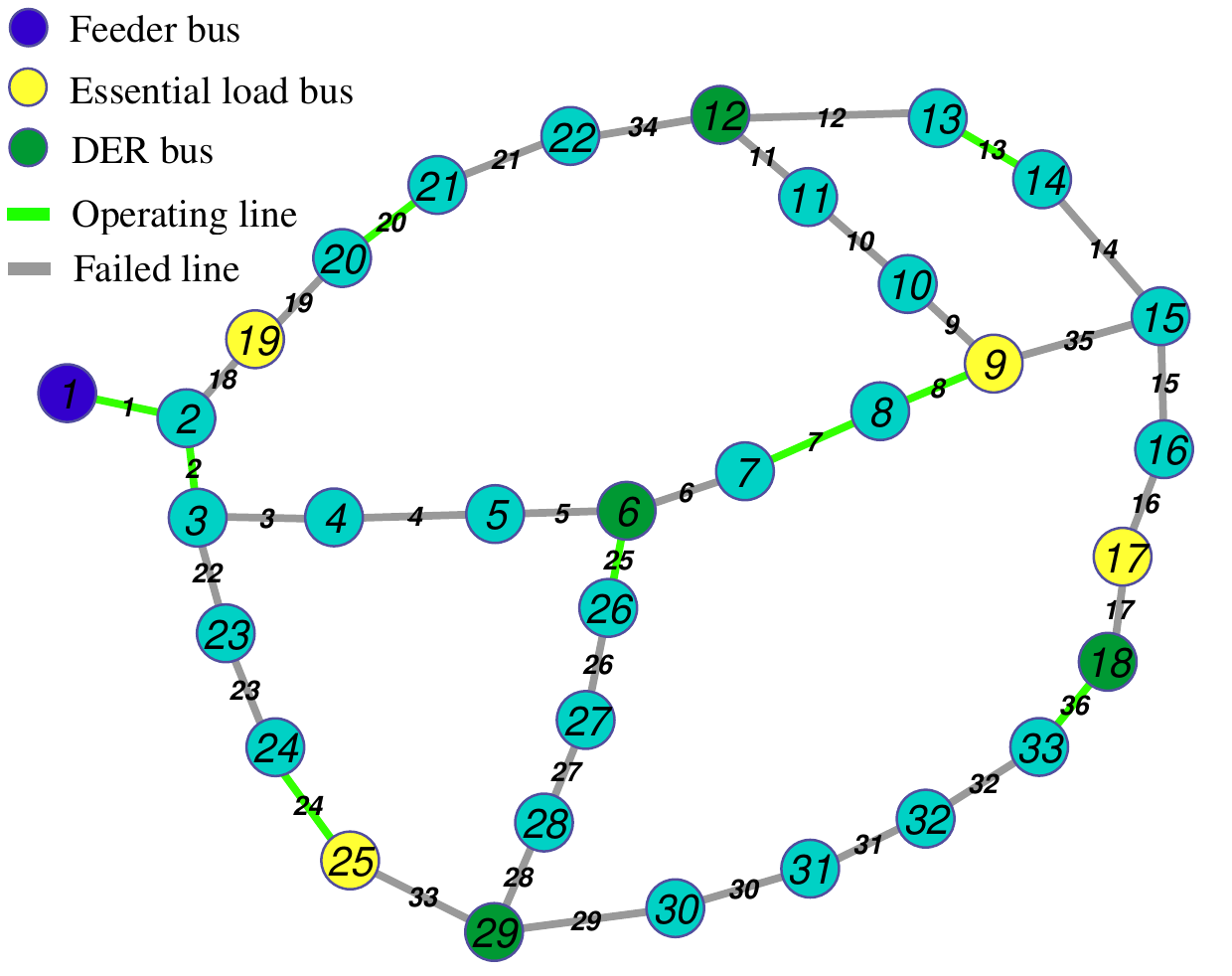}
    \caption{Network graph in one Monte Carlo iteration without hardening.}
    \label{Graph of the network in one of the Monte Carlo iterations without hardening}
\end{figure}

\begin{figure}
    \centering
    \includegraphics[width=\linewidth]{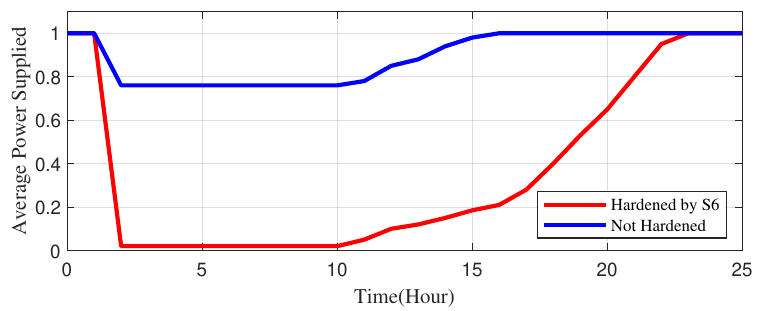}
    \caption{Network resilience Curve before hardening and hardened by S6.}
    \label{Resilience Curve of the network before Hardening and hardened by S6}
\end{figure}

The earthquake is applied to the system at $t=1$. 500 Monte Carlo iterations are executed. Fig. \ref{Graph of the network in one of the Monte Carlo iterations without hardening} shows the graph of the network in one of the Monte Carlo iterations. 
 It can be readily observed from Fig. \ref{Graph of the network in one of the Monte Carlo iterations without hardening} that 25 network lines have been damaged and cut off during the earthquake. In this case, the post-earthquake distribution network is not capable of supplying a large part of the network loads. The average power supply of the network in this case can be seen in Fig. \ref{Resilience Curve of the network before Hardening and hardened by S6} that presents the network resilience curve, i.e., average power supplied.
As seen in Fig. \ref{Resilience Curve of the network before Hardening and hardened by S6}, before the earthquake, all network loads are being supplied. After the earthquake occurs at $t=1$, the level of the supplied load drops sharply and reaches to nearly zero. The cause of this severe power loss is the widespread failures of grid lines and the disconnection between the loads and energy sources. The recovery phase starts at $t=10$ and the disconnected lines will return to orbit every hour. This procedure continues until all lines return to the network, i.e., $t=22$. It is clear that after the recovery phase and the returning of lines to the network, the load supply level gradually increases to $1$. Considering the amount of power supplied at any given moment and the resilience curve, the resilience indices before hardening are $\zeta_V$=0.98, $\zeta_D$=0.95 and, $\zeta_E$=0.87.
It can be readily read that all essential loads are lost after the earthquake happens, indicating poor network resilience. The target criteria are presented in TABLE \ref{goal indxes}.
\begin{table}[!htb]
\centering
\caption{\label{goal indxes} Network Goals and Criteria}
\begin{tabular}{c c c c c}
\hline
    Budget & $\zeta_V$ & $\zeta_D$ & $\zeta_E$ & \% of Essential Load \\
     & & & & Supplied in degraded state\\ [0.5ex]
    \hline
    \$150,000 & $\leq 0.35$ & $\leq 0.35$ & $\leq 0.25$ & $\geq 95\%$ \\
    \hline
\end{tabular}
\end{table}
\subsection{Hardware Hardening Strategies}
By using the proposed strategy generation framework presented in Section \ref{p2s3}, all possible strategies are listed in TABLE \ref{cost benefit list}.
After performing the cost-benefit analysis based on \eqref{benefit}-\eqref{cost to benefit}, all strategies are listed in descending order based on the benefit-to-cost ratio $\alpha(s)$ as shown in TABLE \ref{priority list}. \begin{table}[!htb]
\centering
\caption{\label{cost benefit list}Cost and Benefit Factors for Each Strategy}
\begin{tabular}{c c c c c c}
\hline
    Strategy & Hardened Lines & $C(s)$ & $B(s)$ & $\lambda(s)$ & $\alpha(s)$ \\
    \hline 
    S1 &1,6-8,17,18,33& 75700 & 468819 & 1.42 & 14.863\\
    S2 &1,6-8,12-16,18,33 &115700 & 543454 & 1.83 & 13.292\\
    S3 &6-8,17,19-21,33,34 &89400 & 502358 & 1.64 & 14.834\\
    S4 &1,2,6-8,17,18,22-24 &86000 & 493360 & 1.65 & 15.184\\
    S5 &1,2,6-8,12-18,22-24 &126000 & 583120 & 2.12 & 14.414\\
    S6 &1,2,6-8,17,19-24,34& 109700 & 549339 & 2.01 & 15.050\\
    S7 &1,9-11,17,18,33& 75700 & 465519 & 1.48 & 15.249\\
    S8 &1,9-16,18,33 &115700 & 558579 & 1.95 & 14.239\\
    S9 &9-16,19-21,33,34 &129400 & 579618 & 2.03 & 13.571\\
    S10 &1,2,9-11,17,18,22-24 &109400 & 510619 & 1.72 & 12.694\\
    S11 &1,2,9-16,18 &113200 & 552144 & 1.85 & 13.899\\
    S12 &1,2,9-16,19-24,34 &149700 & 639099 & 2.2 & 13.660\\
\hline
\end{tabular}
\end{table} 

To qualify as an optimal strategy, a hardening strategy must satisfy the goals defined for the network in TABLE \ref{goal indxes}. \begin{table}[!htb]
    \centering
    \caption{\label{priority list}Prioritized List of Strategies}
    \begin{tabular}{c c c c c c}
    \hline
    Priority & Strategy & $\alpha(s)$ &$\zeta_V^s$ & $\zeta_D^s$ & $\zeta_E^s$\\
    \hline
     1 & S7 & 15.249 & 0.56 & 0.53 & 0.47\\
     2 & S4 &15.184 &0.44 &0.38 &0.43 \\
     3 &S6 &15.050 &0.29 &0.28 &0.22\\
     4 &S1 &14.863 &0.45 &0.47 &0.46\\
     5 &S3 &14.834 &0.39 &0.36 &0.41\\
     6 &S5 &14.414 &0.26 &0.23 &0.19\\
     7 &S8 &14.239 &0.34 &0.3 &0.21 \\
     8 &S11 &13.899 &0.38 &0.36 &0.21\\
     9 &S12 &13.660 &0.23 &0.21 &0.17 \\
     10 &S9 &13.571 &0.32 &0.28 &0.18\\
     11 &S2 &13.291 &0.4 &0.36 &0.21 \\
     12 &S10 &12.694 &0.33 &0.33 &0.24\\
     \hline
    \end{tabular}
\end{table}It can be observed from TABLE \ref{priority list} that strategy S7 with the highest benefit-to-cost ratio is at the top. However, its resilience indices cannot meet the targets set in TABLE \ref{goal indxes}. Although it succeeds in supplying all the necessary loads, it performs poorly in strengthening the network resilience -- as high as $\zeta_E^{S7}=47\%$ of the network demanding energy is not provided in the whole process of earthquake occurrence and recovery. For strategy S7, $\zeta_V^{S7}=0.56$ indicates that 56\% of network loads are disconnected at the time of network loss. Therefore, this strategy fails to be the optimal one; same as strategy S4. In contrast, strategy S6 can meet all four targets set in this scenario, thus being the optimal hardening strategy. 

According to TABLE \ref{cost benefit list} and TABLE \ref{priority list}, if the goal was to provide only the essential network loads after the earthquake, then the optimal strategy would be strategy S7. In addition to supplying all essential loads, if the goal was to look for more resilience improvement of the network, then strategy S12 would be the optimal option. In fact, the selection of the optimal strategy depends on the selecting criteria set by the network operator. In this paper, we select a strategy that, along with satisfying all the favorable conditions we have been looking for, has the highest benefit-to-cost ratio -- strategy S6. 

\begin{figure}[!htb]
    \centering
    \includegraphics[width=0.75\linewidth]{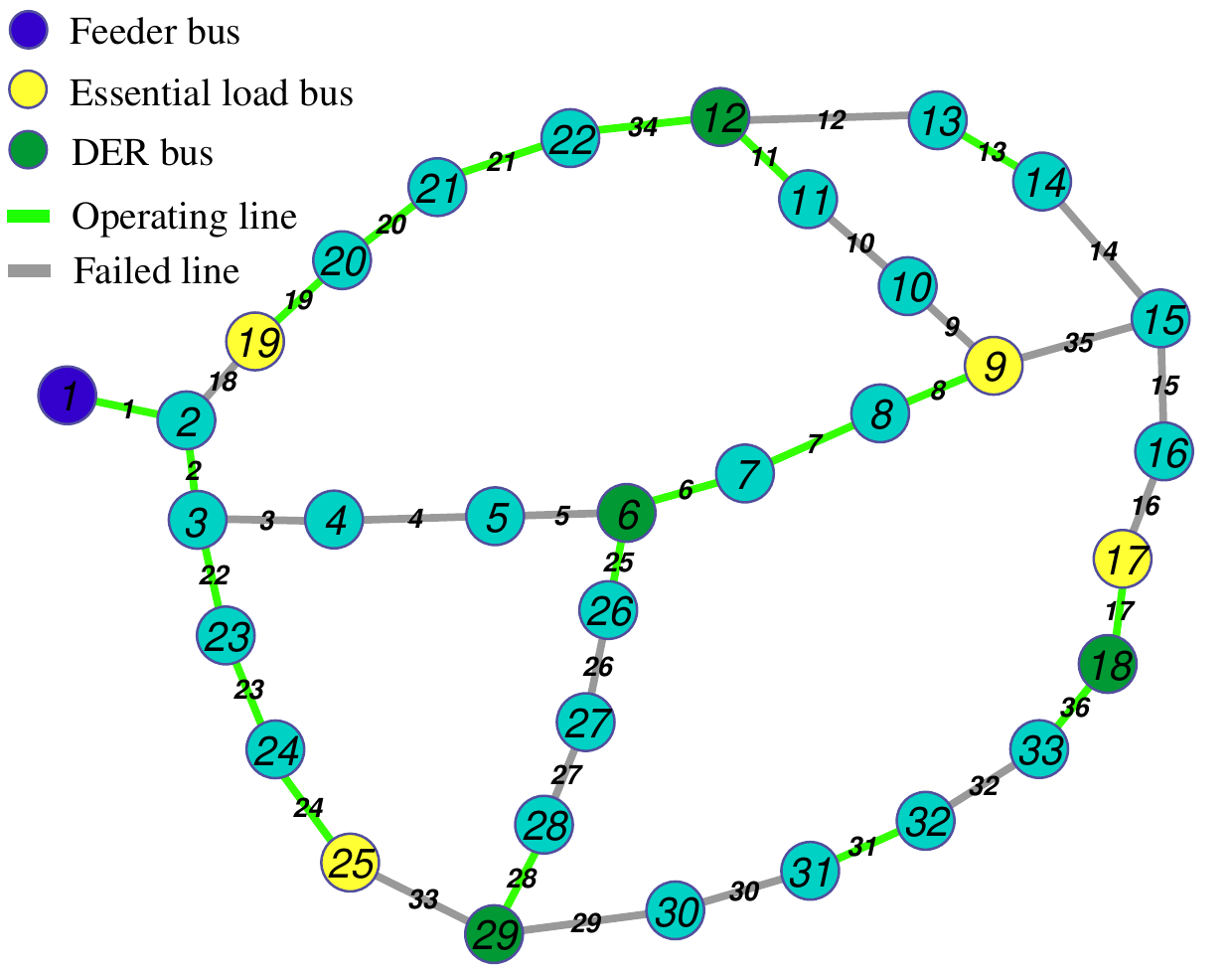}
    \caption{Network graph after adopting S6 in one Monte Carlo iteration.}
    \label{after S6}
\end{figure}

By applying strategy S6 to the IEEE 33-bus network, the graph of the grid in one Monte Carlo iteration is shown in Fig. \ref{after S6}.
It can be readily observed that the lines hardened by this strategy are immune from failure and disconnection. In this case, the network is converted to 15 sub-graphs and only 8 nodes are completely separated from the network. However, all nodes with essential loads are connected to the generation resources by the hardened lines. Admittedly, many lines are damaged from the earthquake, but strategy S6 is able to reduce the cost of cutting off all the essential loads.

The network resilience curve after hardening by using strategy S6 is shown in Fig. \ref{Resilience Curve of the network before Hardening and hardened by S6}. Comparing with the non-hardened case, the hardening leads to an increase in the level of $R_{pe}$, so that, according to TABLE \ref{priority list}, about $1-\zeta_V^{S6}=71\%$ of the total load of the network is supplied. In addition, since the number of damaged lines is reduced, the network recovery process is finished in shorter time. This significant reduction in network recovery time reduces unsupplied energy during accident: In the case of no hardening, according to calculated indices before hardening and TABLE \ref{priority list} only $1-\zeta_E=13\%$ of subscribers' energy is supplied; this amount has reached $1-\zeta_E^{S6}=88\%$ after hardening. By implementing S6, 100\% of the essential loads can be supplied in the degraded phase. According to TABLE \ref{priority list}, the resilience improvement $\lambda(S6)$ of strategy S6 is $2.12$, indicating that strategy S6 has the optimal performance \emph{w.r.t.} all three indices.
\section{Conclusion}
In this paper, a hardening-oriented framework was developed to improve the distribution network resilience in the face of earthquake. We introduced three normalized resilience indices to better evaluate the network resilience during and after the earthquake. A comprehensive framework was developed to determine the optimal hardware hardening strategy with the consideration of resilience improvement and economy. Simulation results on the IEEE 33-bus test feeder showed that implementing the optimal hardening strategy significantly improves the distribution network resilience and the hardened network can supply all essential loads after the earthquake happens. This paper has successfully showed that hardening the network by using the proposed method can dramatically improve the distribution network resilience, lower the possible loss of the network, and prevent significant economic loss. 

\bibliographystyle{IEEEtran}

\bibliography{conference_101719}

\end{document}